\documentstyle[epsfig,aps,prd,twocolumn]{revtex}

\def\etal{{\it et al.}}

\def\~{{$\tilde{\phantom{a}}$}}

\begin{document}

\title{\vspace{-0.5in}
Methods of Calculating Forces on Rigid Magnetic Media}

\author{Kirk T.~McDonald}
\address{Joseph Henry Laboratories,
Princeton University, Princeton, New Jersey 08544}

\date{March 18, 2002} 

\maketitle

\begin{abstract}
Despite 180 years of theory on magnetism, it appears that the practice of
calculating forces on magnetic media is ambiguous, as illustrated by a recent
article in this Journal \cite{Casperson}.  Potentially troubling issues include:
Which field {\bf B} or {\bf H} should be used?  
Should the total field be used, or only the external field?  
And if the latter, what is meant by ``external''?
Can/should effects of magnetization currents and/or fictitious
magnetic poles be included?
What is the force on only a portion of a circuit?
We review several well-known approaches
to magnetic force calculations on elements of rigid circuits, 
and find it very helpful to use an explicit
example to compare and contrast the various methods.  Our discussion reinforces
that of the most authoritative texts \cite{Stratton,Landau}, but corrects in an
important way a previous attempt at a systematic review in this Journal
\cite{Brown1,Brown2,Brown3,Brown4}.
\end{abstract}  

\section{Introduction} 

The phenomenon of magnetism was first manifested via interactions of bulk
magnetic materials (magnets).  Following the discovery by Oersted \cite{Oersted}
that pairs of currents exerted forces on one another, Biot and Savart
\cite{Biot} identified a corresponding force law, and Amp\`ere \cite{Ampere}
made the conjecture that
all magnetic phenomena are actually due to currents, some of which may be
bound in ``molecules'' of magnetic materials.  This view contrasted with the
elegant work of Poisson \cite{Poisson} in which forces on magnetic materials
(without conduction currents) could be explained in terms of a density of 
magnetic poles bound in the media in such a way that single poles can never be
isolated.   

Pedagogic discussions of forces on magnetic media tend to treat the case of
current-carrying conductors separately from that of materials (typically
conductors) that have nonzero bulk magnetization.  The present article 
addresses the general case of magnetic media in which conduction currents
are flowing, but is restricted to the case of forces on elements of a rigid
circuit.  The literature on this topic appears to be both sparse and
erratic, and we have the (possibly unrealistic) hope of clarifying the record.

Because the details prove to be intricate, the best impression that a reader may
form is simply that one must approach magnetic force problems with caution.

Some more specific conclusions from this effort are:
\begin{enumerate}
\item
The consequence of Newton's 1st law that an object does not exert a net force on
itself is a guiding principle in the analysis of static forces on rigid circuits.
See sec.~III.
We do not pursue here the more complex case of deformable magnetic media.
\item
In calculating the force on a circuit (or on a portion of a rigid circuit) one
can use only that magnetic field which existed prior to the introduction of the
circuit, since the fields produced directly (or indirectly via induced 
magnetization) cannot cause a net force on the circuit (or portion of a rigid
circuit).  See sec.~III.
\item
Methods also exist in which the total magnetic field can be used in the force
calculation, even if the force on only a portion of a rigid circuit is desired.
The most straightforward of these uses the Maxwell stress tensor
(sec.~VIII),
which the author recommends be given more prominence in pedagogic 
treatments of electrostatic and magnetostatic forces.
\item
Considerable care must be given in the treatment of systems that include an
interface across which the magnetic permeability changes discontinuously.
See secs.~IV-IX.
\item
If the Helmholtz bulk force density (calculated from the total magnetic fields)
is used to find the force by integration over a volume that contains
such an interface, the volume integral must be supplemented by a surface
integral over the (interior) interface.  See sec.~IX.
\item
In a calculation of the force that uses the concept of magnetization currents, 
the magnetic field {\bf H} should be used rather than the field {\bf B}, and if
the force is desired on only a circuit element, the field must be ${\bf H}_i$ as
existed prior to the introduction of the circuit (and not merely the introduction
of the circuit element). See sec.~VI.
\item
In a calculation of the force that uses the concept of magnetic poles, the
magnetic field {\bf B} should be used rather than the field {\bf H}, and the
prior field ${\bf B}_i$ should be used if the force on only a circuit element is
desired.  See sec.~VII.
\end{enumerate}   

While conclusions 5 and 6 may be known to those expert with magnetic media,
the author has not found them in the literature.   Because several of the
conclusions amend conventional pedagogic wisdom, it has been critical to
verify these conclusions with an example in
which both ${\bf I} \times {\bf B}$ forces and magnetization forces are present,
and which contains an interface with different permeabilities on either side.                        

\section{An Example}

What is the force per unit length on a wire of radius $a$ and (relative) 
permeability $\mu'$ when it carries uniform conduction 
current density 
\begin{equation}
{\bf J}_{\rm cond} = {I \over \pi a^2} \hat{\bf z} 
\label{p00}
\end{equation}
and is placed along the $z$ axis in a magnetic field whose form is 
\begin{equation}
{\bf B}_i = B_0 \hat{\bf x} + B_1 \left[ {x \over a} \hat{\bf x} - {y \over a} \hat{\bf y}
\right]
\label{p0}
\end{equation} 
before the wire is placed in that field?
The medium surrounding the wire is a liquid with relative permeability $\mu \ne 1$.
The wire is assumed to be part of a rigid circuit that is completely by a loop at
``infinity''.  

Note that in asking for the force on the wire along the $z$ axis,
we seek the force on only a portion of the complete circuit. 
    
The form of the initial magnetic field has been chosen so that there will 
be both a ${\bf J} \times {\bf B}$ force associated with the uniform field 
${\bf B}_0 = B_0 \hat{\bf x}$, 
as well as a force due to the interaction of the induced magnetization with the
nonuniform field ${\bf B}_1 = B_1 [(x / a) \hat{\bf x} - (y / a) \hat{\bf y}]$.

We have confirmed by the four methods of calculation discussed
in secs.~VI-IX that the force per unit length on the wire is, in Gaussian units,
\begin{equation}
{\bf F} = - {\mu - \mu' \over \mu + \mu'} {a B_0 H_1 \over 2} \hat{\bf x}
+ {I B_0 \over c} \hat{\bf y},
\label{p1}
\end{equation}
where the magnetic fields {\bf H} and
\begin{equation}
{\bf B} = {\bf H} + 4 \pi {\bf M},
\label{p2}
\end{equation}
are related in a linear, isotropic
medium of relative permeability $\mu$ by
\begin{equation}
{\bf B} = \mu {\bf H},
\label{p3}
\end{equation}
and hence the magnetization density {\bf M} obeys
\begin{equation}
{\bf M} = {\mu - 1 \over 4 \pi} {\bf H}.
\label{p4}
\end{equation}
The magnetic field {\bf H} for this example is given in the Appendix, and
details of the calculations can be found in \cite{permeable-wire}.

If the initial magnetic field is uniform ($B_1 = 0$), then there is no net force
on the magnetization of the wire or surrounding medium, and the force per unit
length on the wire has the very simple form 
\begin{equation}
{\bf F} = {I B_0 \over c} \hat{\bf y}  \qquad \mbox{(uniform\ external\ field)}.
\label{p5}
\end{equation}

The expression (\ref{p5}) has been verified in a recent experiment Casperson
\cite{Casperson}.  See also a related discussion by Lowes \cite{Lowes}.

Strictly speaking, eq.~(\ref{p5}) describes
 the force on the conduction electrons, and not that
on the lattice of positive ions through which the electrons flow.  The force
(\ref{p5}) results in a slight rearrangement of the
distribution of the conduction electrons and positive lattice ions so that
a transverse electric field is generated that acts on the lattice to provide the
force experienced by an observer who holds the wire at rest.   See \cite{McKinnon}
for further discussion.

\section{The Biot-Savart Force Law}

The result (\ref{p5}) is to be expected from a simple argument, consistent with Newton's
first law that an object does not exert a net force on itself.  Namely, the
Biot-Savart force law (for media of unit permeability) 
on a current-carrying circuit $a$ due to a current-carrying
circuit $b$ is
\begin{equation}
{\bf F}_a = {I_a I_b \over c^2} \oint_a d{\bf l}_a \times \oint_b {d{\bf l}_b \times
\hat{\bf r}_{ab} \over r_{ab}^2}
= {I_a \over c} \oint_a d{\bf l}_a \times {\bf B}_b(a),
\label{p6}
\end{equation}
where
\begin{equation}
{\bf B}_b(a) = {I_b \over c} \oint_b {d{\bf l}_b \times
\hat{\bf r}_{ab} \over r_{ab}^2}\, .
\label{p7}
\end{equation}

In the example of sec.~II, the initial field ${\bf B}_i$ plays the role of the
field ${\bf B}_b$ not due the current in wire $a$, which quickly leads to the
result (\ref{p5}).
The reader may object that the example does not involve media of unit
permeability, so the Biot-Savart force law may have to be modified in such a way
as to lead to a different result than (\ref{p5}).  This  issue will be
pursued in the following section.

The statement that an object cannot exert a net force on itself tacitly
presumes that the
object has mechanical integrity as a whole, and will generate internal elastic
forces to counteract possible electromagnetic forces of one part of the object on
another.  If a magnetic circuit is mechanically flexible, we will not be
content with an analysis of the force on the circuit as a whole.  In addition,
we desire a calculation of magnetic force on an element of the circuit,
imagining it to be mechanically
(but not electrically) disconnected from the rest of the circuit.  In this case,
the magnetic field to be used in differential form to the Biot-Savart force law
is the field ${\bf B}_{\rm ext}$ due to all sources outside the element itself,
\begin{equation}
d{\bf F} = {I \over c} d{\bf l} \times {\bf B}_{\rm ext}.
\label{p6a}
\end{equation}
The differential expression (\ref{p6a}) can have meaning independent of the 
integral form (\ref{p6}) only if the circuit deforms, in which case the
problem is not one of statics.  It is well known that the differential force 
(\ref{p6a}) does not satisfy Newton's 3rd law in cases of isolated current
elements, because in such dynamical systems the
electromagnetic field momentum is varying \cite{Page}.  
Newton's 3rd law is respected
via the appropriate electromagnetic version of his 2nd law,
\begin{equation}
\sum \left( {\bf F} - {d{\bf P}_{\rm mech} \over dt} - {d{\bf P}_{\rm EM} \over dt}
\right) = 0,
\label{p6b}
\end{equation}
where ${\bf P}_{\rm mech}$ and ${\bf P}_{\rm EM}$ are the mechanical and 
electromagnetic momenta, respectively.   For a discussion of deformable
circuits, see \cite{Cavalleri}.  The topic of forces on magnetic liquids,
including striction effects, is treated in \cite{Rosensweig}.

In the remainder of this article we assume that the object on which we desire
to calculate the magnetic force is a rigid body at rest.

\section{Microscopic and Macroscopic Magnetic Fields}

In the macroscopic approach to calculation of forces on magnetic media one
considers in general a current density {\bf J} (electric charge crossing a
directed unit area per unit time) and a net magnetization density
{\bf M} (net magnetic dipole moment per unit volume) as well as the
magnetic fields {\bf B} and ${\bf H} = {\bf B} - 4 \pi {\bf M}$. 
In a microscopic view
one considers collections of individual charges and/or magnetic moments, but
averaged quantities like the magnetization density {\bf M} are not yet
defined.  Hence,
the fields {\bf B} and {\bf H} are identical in the microworld, and it is a matter
of convention which symbol is used for the microscopic magnetic field.

The symbol {\bf H} was used for the magnetic field by the early workers
Amp\`ere, Biot and Savart, and Poisson \cite{Poisson}, which led authors such as Lorentz \cite{Lorentz} and Landau \cite{Landau}
 to use this symbol for the microscopic magnetic field.  The symbol {\bf B}
was introduced by Thomson around 1850 \cite{Thomson} in the form (\ref{p3}),
which suggests that it is to be derived from the more fundamental (or anyway
more familiar at the time) field {\bf H}.  However, as
apparently first noted by Lorentz \cite{Lorentz}, the macroscopic average of the 
microscopic magnetic field is {\bf B} and not {\bf H}.

To maximize the continuity between the microscopic and 
macroscopic views, the author prefers that the microscopic magnetic field be
labeled {\bf B}.  Then, the force on an electric charge $e$
with velocity {\bf v} in microscopic electromagnetic fields {\bf E} and {\bf B} is
\begin{equation}
{\bf F} = e \left( {\bf E} + {{\bf v} \over c} \times {\bf B} \right),
\label{s2.1}
\end{equation}
the Lorentz force law, which also represents the average force on the charge in
the presence of macroscopically averaged fields {\bf E} and {\bf B}.
It is always understood that the fields {\bf E} and {\bf B} in the Lorentz force
law do not include the fields of the moving charge itself.

High-energy physicists such as the present author, consider the Lorentz force
law (\ref{s2.1}) to be
experimentally confirmed (and continually reconfirmed) for over 50 years 
\cite{Rasetti} in the case of high-energy particles moving inside media such as
iron where ${\bf B} = \mu {\bf H} \gg {\bf H}$.  That is, when applying the
Lorentz force law a charge inside a macroscopic medium, the appropriate
macroscopic average of the microscopic magnetic field is indeed the macroscopic field
{\bf B}.  This insight is affirmed in sec.~8.2 of \cite{Panofsky}, and
sec.~22.1.1 of \cite{Corson},

We therefore expect that other methods of calculating forces on currents in
macroscopic media will be consistent with the Lorentz force law using the
macroscopic field {\bf B}.  In particular, we expect that the Biot-Savart force
law for a current {\bf I} in a macroscopic magnetic field ${\bf B}_i$ not due to
that current itself is
\begin{equation}
{\bf F} = {1 \over c} \oint {\bf I} dl \times {\bf B}_i
= {I \over c} \oint  d{\bf l} \times {\bf B}_i,
\label{s2.2}
\end{equation}
no matter what permeabilities exist.  Thus, we reaffirm that eq.~(\ref{p5}) is the
force on the wire in our example, if the initial field is uniform.

\section{The Biot-Savart Force Law in a Permeable Medium}

Despite the simplicity of the result (\ref{s2.2}), care is needed when using the 
Biot-Savart force
law in permeable media. We review this issue by starting with the  case that 
all wires and their
surrounding media have the same permeability $\mu \neq 1$.  Then there is neither a
surface current nor a fictitious pole density at the interface between the wire
and the liquid.  However, there remains a volume current density 
\begin{equation}
{\bf J}_M = c \nabla \times {\bf M} 
= {\mu - 1 \over 4 \pi / c} \nabla \times {\bf H}
= (\mu - 1) {\bf J}_{\rm cond},
\label{s3.1}
\end{equation}
using Amp\`ere's law,
\begin{equation}
\nabla \times {\bf H} = {4 \pi \over c} {\bf J}_{\rm cond},
\label{s3.2}
\end{equation}
that relates the magnetic field {\bf H} to the conduction current density 
${\bf J}_{\rm cond}$.  Thus, the total current density is 
\begin{equation}
{\bf J}_{\rm total} = \mu {\bf J}_{\rm cond}.
\label{s3.3}
\end{equation}

The fact that the total current density (\ref{s3.3}) does not equal the conduction current
density in a permeable medium contradicts the view of Lorentz \cite{Lorentz},
as reaffirmed in sec.~30 of \cite{Landau}.  This appears to be one of the very
few oversights in these distinguished works.

The force (\ref{s2.2}) on the wire that carries conduction current density 
${\bf J}_{\rm cond}$ can be written as 
\begin{equation}
{\bf F} = {1 \over c} \int {\bf J}_{\rm cond} \times {\bf B}_i \ d{\rm Vol}.
\label{s201}
\end{equation}
If instead we wish to use the total current density (\ref{s3.3}) we must write
\begin{equation}
{\bf F} = {1 \over c} \int {{\bf J}_{\rm total} \over \mu} \times {\bf B}_i \ d{\rm Vol}
= {1 \over c} \int {\bf J}_{\rm total} \times {\bf H}_i \ d{\rm Vol}.
\label{s202}
\end{equation}


Another aspect of the analysis of Biot and Savart is the calculation of the magnetic
field from the current density.  The microscopic version of Amp\`ere's law,
\begin{equation}
\nabla \times {\bf B} = {4 \pi \over c} {\bf J}_{\rm total},
\label{s203}
\end{equation}
corresponds to the prescription that
\begin{equation}
{\bf B} = {1 \over c} \int {{\bf J}_{\rm total} \times \hat{\bf r} \over r^2} d{\rm Vol}
= {\mu \over c} \int {{\bf J}_{\rm cond} \times \hat{\bf r} \over r^2} d{\rm Vol}
= \mu {\bf H},
\label{s204}
\end{equation}
supposing the permeability is uniform and may be taken outside the integral.
Hence, the macroscopic version of Amp\`ere's law, eq.~(\ref{s3.2}),
corresponds to the prescription that
\begin{equation}
{\bf H} = {1\over c} \int {{\bf J}_{\rm cond} \times \hat{\bf r} \over r^2} d{\rm Vol}
= \nabla \times {1\over c} \int {{\bf J}_{\rm cond} \over r} d{\rm Vol},
\label{s206}
\end{equation}
independent of the permeability.  This result, combined with Amp\`ere's law (\ref{s3.2}),
 is consistent with Helmholtz' theorem 
\cite{Panofsky} provided $\nabla \cdot {\bf H} = 0$, as holds within a medium of 
uniform permeability.

The form of the eq.~(\ref{p6}) for the force on circuit $a$ due to circuit $b$
supposing the wires and the surrounding media all have permeability $\mu$ is therefore
\begin{equation}
{\bf F}_a  
= \mu {I_a I_b \over c^2} \oint_a d{\bf l}_a \times \oint_b {d{\bf l}_b \times
\hat{\bf r}_{ab} \over r_{ab}^2}\, ,
\label{s207}
\end{equation}
where $I_a$ and $I_b$ are the conduction currents in the circuits.  

\section{The Biot-Savart Force Law Plus Bound Current Densities}

We also see that eq.~(\ref{s207}) holds even if
the wires have permeabilities $\mu_a$ and $\mu_b$ that differ from the permeability $\mu$
of the surrounding medium, since the magnetic field due to wire $b$ at the position of wire 
$a$ before wire $a$ was introduced is given by ${\bf B}_b = \mu {\bf H}_b$, which
depends on neither $\mu_a$ nor $\mu_b$.  However, in this case there will exist
effective surface currents, 
\begin{equation}
{\bf K}_M = c \Delta {\bf M} \times \hat{\bf n},
\label{s3.4}
\end{equation}
at the interface between the wires and the surrounding medium, where 
$\Delta {\bf M}$ is the difference between the magnetization on the two sides of the
interface, and $\hat{\bf n}$ is
the unit normal to the interface.  The force on these surface currents follows the
form (\ref{s202}) that uses the magnetic field {\bf H} rather than {\bf B}.  Hence
the total force on a permeable wire surrounded by a permeable medium can be
written as
\begin{equation}
{\bf F} = {1 \over c} \int  {\bf J}_{\rm total} \times {\bf H}_i\ d{\rm Vol}
+ {1 \over c} \int  {\bf K}_M \times {\bf H}_i\ dS.
\label{s3.5}
\end{equation}
The is the appropriate version of the Biot-Savart law if we wish to include 
magnetization forces via the so-called bound current densities.  
However, the Coulomb Committee in their eq.~(1.3-4$'$) \cite{Brown3},
and Jefimenko in his eq.~(14-9.13a,b) \cite{Jefimenko},
recommends that the initial field ${\bf B}_i$ be used rather
than ${\bf H}_i$, which would imply a force $\mu$ times the above.

For the example of sec.~II, the surface current density is obtained from eq.~(\ref{p4})
and the total magnetic field {\bf H}, eq.~(\ref{s16}), as
\begin{equation}
{\bf K}_M = {\mu - \mu' \over 4 \pi} \left[ {2 I \over a} 
- {2 \mu c \over \mu + \mu'}  (H_0 \sin\theta + H_1 \sin 2\theta) \right] \hat{\bf z},
\label{s3.6}
\end{equation}
and the total current density is ${\bf J}_{\rm total} = \mu' I \hat{\bf z} / \pi a^2$ from
eqs.~(\ref{p00}) and (\ref{s3.3}).  Then, evaluation of eq.~(\ref{s3.5}) leads to 
eq.~(\ref{p1}).

The correct result for the force on a portion of a rigid circuit
would not be obtained from eq.~(\ref{s3.5}) if we used the initial
magnetic field ${\bf B}_i$, or if we used the total
magnetic fields {\bf B} or {\bf H} on the wire.  Furthermore, we would not obtain a correct
result if we used the field in a vacuum cavity of radius $a$ at the position of the wire
either before or after the wire was inserted.  The proper initial field ${\bf H}_i$ is the one
before the wire was inserted into the liquid dielectric.

When $\mu \ne \mu' \ne 1$, we cannot rewrite the first term of eq.~(\ref{s3.5}) as
$\int  ({\bf J}_{\rm cond} / c) \times {\bf B}_i\ d{\rm Vol}$,
which would incorrectly suggest that we could ignore the volume magnetization current
density ${\bf J}_M$ but not the surface current density ${\bf K}_M$.

\section{The Biot-Savart Force Law Plus Fictitious Magnetic Poles}

Following Poisson \cite{Poisson},
the forces on the magnetization of the media can also considered as due to a
density of fictitious magnetic poles,  rather than being due to currents ${\bf J}_M$
and ${\bf K}_M$.  Some care is required to use this approach,
since a true magnetic pole density $\rho_M$ would imply $\nabla \cdot {\bf B}
= 4 \pi \rho_M$, and the bulk force density on these poles
would be ${\bf F} = \rho_M {\bf B}$.  However, in reality $0 = \nabla \cdot {\bf B} 
= \nabla \cdot ({\bf H} + 4 \pi {\bf M})$,
so we write 
\begin{equation}
\nabla \cdot {\bf H} = - 4 \pi \nabla \cdot {\bf M} = 4 \pi \rho_M,
\label{6.1}
\end{equation}
and we identify $\rho_M = - \nabla \cdot {\bf M}$ as the volume density of 
fictitious magnetic poles.
Inside linear magnetic media, such as those considered here, ${\bf B} = \mu
{\bf H}$ and $\nabla \cdot {\bf B} = 0$ together imply that $\rho_M = 0$.
However, a surface density $\sigma_M$ of fictitious poles
can exist on an interface between two media, and we see that Gauss' law for the field
{\bf H} implies that
\begin{equation}
\sigma_M = { ({\bf H}_2 - {\bf H}_1) \cdot \hat{\bf n} \over 4 \pi}\, ,
\label{6.2}
\end{equation}
where unit normal $\hat{\bf n}$ points across the interface from medium 1 to medium 2.
The surface pole density can also be written in terms of the
magnetization ${\bf M} = ({\bf B} - {\bf H}) / 4 \pi$ as
\begin{equation}
\sigma_M = ({\bf M}_1 - {\bf M}_2) \cdot \hat{\bf n},
\label{6.3}
\end{equation}
since $\nabla \cdot {\bf B} = 0$ insures that the normal component of {\bf B} is
continuous at the interface.

The force on the surface density of fictitious magnetic poles is
\begin{equation}
{\bf F} = \sigma_M {\bf B}_i,
\label{6.4}
\end{equation}
since the fictitious poles couple to the macroscopic average of the microscopic
magnetic field, as anticipated by Thomson and Maxwell \cite{BorH}.
Equation (\ref{6.4}) is in agreement with prob.~5.20 of \cite{Jackson}.
However, the Coulomb Committee in their eq.~(1.3-4) \cite{Brown3},
and Jefimenko in his eq.~(14-9.9a,b) \cite{Jefimenko},
recommends that the initial field ${\bf H}_i$ be used rather
than ${\bf B}_i$ when using the method of fictitious magnetic poles, which
would imply a force $1 / \mu$ times the above.  

The total force on the medium in this view is the sum of the force on the
conduction current plus the force on the fictitious surface poles, where to avoid calculating
a spurious force of the rigid
wire on itself we use the initial magnetic field ${\bf B}_i$,
\begin{equation}
{\bf F} =  {1 \over c} \int {\bf J}_{\rm cond} \times {\bf B}_i\ d{\rm Vol}
+  \int \sigma_M {\bf B}_i\ dS.
\label{6.5}
\end{equation} 

In the example of sec.~II, the density of fictitious magnetic poles on the surface
$r = a$ is given by
\begin{eqnarray}
\sigma_M & = & { H_r(r=a^+) - H_r(r=a^-) \over 4 \pi}
\nonumber \\
& = & - {1 \over 2 \pi} {\mu - \mu' \over \mu + \mu'} (H_0 \cos\theta + H_1 \cos 2\theta).
\label{6.6}
\end{eqnarray}
Then, evaluation of eq.~(\ref{6.5}) again leads to the result (\ref{p1}).
This reaffirms that one should use the initial field ${\bf B}_i$ and not ${\bf H}_i$
and not the total field {\bf B} in the Biot-Savart method (\ref{6.5})
with fictitious magnetic poles when calculating the force on only a portion of
 a rigid circuit.

\section{The Maxwell Stress Tensor}

The methods of calculating the force on a circuit discussed thus far are require care 
in that only the fields ${\bf B}_i$ and ${\bf H}_i$ prior to the addition of the 
circuit appear explicitly in the force calculation, although the total fields {\bf B}
and {\bf H} are needed in the calculation of the magnetization.
The erratic literature on this topic is ample evidence that confusion as how to
implement these calculation is likely.

Hence, it may be preferable to use methods that involve only the total fields {\bf B}
and {\bf H}.  In the author's view, the most reliable general method for the 
calculation of electromagnetic forces is that based on the Maxwell stress tensor
\cite{Maxwell641}, which is a formal transcription of Faraday's ``tubes of force''.
The $j$ component of the electromagnetic force {\bf F} on the interior of a 
closed volume in a linear medium with oriented surface element $d{\bf S}$ is given by
\begin{equation}
F_j = \int \sum_k T_{jk} dS_k,
\label{8.0}
\end{equation}
where
\begin{equation}
T_{jk} 
= {1 \over 4 \pi}  E_j D_k + B_j H_k - {\delta_{jk} \over 8 \pi} 
( {\bf E} \cdot {\bf D} + {\bf B} \cdot {\bf H}).
\label{8.1}
\end{equation}
The form (\ref{8.1})
ignores interesting strictive effects in compressible media \cite{Brevik}.

The surface over which the stress tensor is integrated need not correspond to a
physical surface, which leads to the question of how the electromagnetic force
is transmitted to the physical matter inside that surface.  The answer is,
of course, via the electromagnetic fields that enter into the stress tensor.
Prior to, and even somewhat after, Maxwell, more physical explanations were 
considered necessary, which led to the diversionary search for the {\ae}ther
that, among other activities,
would transmit the forces from the imaginary surface to the matter within.

We have already noted in sec.~II that even the ${\bf J} \times {\bf B}$ force
concept is at least once removed from a force on the positive ion lattice of
a conductor -- which caused confusion to Maxwell \cite{Maxwell501}.

For the example of sec.~II, we 
calculate the force on unit length of the wire by integrating the Maxwell stress 
tensor over a cylindrical surface of radius $r > a$,
so that any effects at the surface $r = a$ are included.  
The result is, as expected, given by eq.~(\ref{p1}), independent of radius $r$.

Equation (\ref{p1}) for the case that $B_1 = 0$ was deduced in a similar manner in 
ref.~\cite{Stratton}. 

If we integrate the stress tensor over
 a cylinder of radius $r < a$ the result is
\begin{equation}
{\bf F} = 
{2 \mu' \over \mu + \mu'} {I B_0 \over c} {r^2 \over a^2.} \hat{\bf y}.
\label{8.2}
\end{equation}
Since the limit of this as
$r \to a$ does not equal the result (\ref{p1})
for $r > a$, we infer that there are important effects at the interface
$r = a$.  The permeable liquid is presumably contained in a tank of some
characteristic radial scale $b \gg a$, at whose surface additional magnetization
forces will arise.  We consider these forces as distinct from those at the 
interface $r = a$, and that only the latter are part of the forces on the wire.

\section{The Helmholtz Bulk Force Density}

An expression for a bulk force density {\bf f} in magnetic media can be
obtained by transformation of the surface integral of the stress tensor into a
volume integral.  See, for example, secs.~15 and 35 of \cite{Landau}.  The result,
again ignoring magnetostriction, is
\begin{equation}
{\bf f} = {1\over c} {\bf J}_{\rm cond} \times {\bf B} - {H^2 \over 8 \pi}
\nabla \mu,
\label{s101}
\end{equation}
which is due to Helmholtz \cite{Helmholtz}.  As for the Maxwell stress tensor,
the fields {\bf B} and {\bf H} in eq.~(\ref{s101}) are the total fields from 
all sources.

However, eq.~(\ref{s101}) is not sufficient for the case that the permeability takes
a discontinuous step at an interface within the volume of interest.
To see this, we recall the usual derivation of the bulk force density, beginning with
Gauss' law to transform the surface
integral of the Maxwell stress tensor into a volume integral of a force density {\bf f},
\begin{equation}
F_i = \int \sum_j T_{ij}\ dS_j 
= \int \sum_j {\partial T_{ij} \over \partial x_j} d{\rm Vol}
= \int f_i\ d{\rm Vol}.
\label{s105}
\end{equation}
Recalling eq.~(\ref{8.1}), the force density {\bf f} is given by
\begin{eqnarray}
{\bf f} & = & \nabla \cdot {\bf T}
= {1 \over 4 \pi} \left[ ({\bf B} \cdot \nabla) {\bf H} + {\bf H} (\nabla \cdot {\bf B})
- {1 \over 2} \nabla ({\bf B} \cdot {\bf H}) \right]
\nonumber \\
& = & {\mu \over 4 \pi} ({\bf H} \cdot \nabla) {\bf H}
- {1 \over 8 \pi} H^2 \nabla \mu - {\mu \over 8 \pi}  \nabla H^2,
\label{s106}
\end{eqnarray}
since $\nabla \cdot {\bf B} = 0$ always, and we first suppose that
${\bf B} = \mu {\bf H}$ involves a continuously varying permeability.
The usual argument then proceeds by noting that 
\begin{eqnarray}
\nabla H^2 & = & 2 ({\bf H} \cdot \nabla) {\bf H} + 2 {\bf H} \times (\nabla \times {\bf H})
\nonumber \\
& = & 2 ({\bf H} \cdot \nabla) {\bf H} - {8 \pi \over c} {\bf J}_{\rm cond} \times {\bf H},
\label{s107}
\end{eqnarray}
using Amp\`ere's law (\ref{s3.2}).  
Inserting eq.~(\ref{s107}) in (\ref{s106}) we arrive at 
eq.~(\ref{s101}).  However, if the volume of interest includes an interior interface across
which the permeability takes a discontinuous step, we should revert to the first form of
eq.~(\ref{s106}) when performing the volume integral across the interface.  Defining
$\hat{\bf n}$ to be the unit normal to the interface, and noting that ${\bf B} \cdot {\bf H}
= \mu H_t^2 + B_n^2 / \mu$ where the tangential and normal components, $H_t$ and $B_n$,
are continuous across such an interface, 
the resulting surface integral is
 \begin{eqnarray}
{1 \over 4 \pi} \int \left[ B_n \Delta H_n 
- {1 \over 2} \Delta ({\bf B} \cdot {\bf H}) \right] \hat{\bf n}\, dS,
\nonumber \\
= {1 \over 8 \pi} \Delta\left( {1 \over \mu} \right)  \int B_n^2 \hat{\bf n}\, dS
- {\Delta \mu \over 8 \pi} \int H_t^2 \hat{\bf n}\, dS 
\label{s107a}
\end{eqnarray}
The total magnetic force acan now be written
 \begin{eqnarray}
{\bf F} & = & \int {1 \over c} {\bf J}_{\rm cond} \times {\bf B}\ d{\rm Vol}
- {1 \over 8 \pi} \int H^2 \nabla \mu \ d{\rm Vol}
\nonumber \\
& &  +\ {1 \over 8 \pi} \Delta\left( {1 \over \mu} \right)  \int B_n^2 \hat{\bf n}\, dS
- {\Delta \mu \over 8 \pi} \int H_t^2 \hat{\bf n}\, dS 
\label{s108}
\end{eqnarray}
where the surface integral is over the interface which is, in general, interior to the
volume of integration, and
$\Delta A$ is the difference of quantity $A$ on the two sides of the interface.
To apply eq.~(\ref{s108}) to example of sec.~II, we note that the second integral is zero
(considering the integral over the interior interface as distinct from the bulk
volume integrals),
and that $\hat{\bf n} = \hat{\bf r}$ on the interface at $r = a$.  Therefore, we need
only those terms of $B_r^2$ and $H_\theta^2$ that vary as
$\sin\theta$ or $\cos\theta$. Referring to the Appendix, we find
\begin{eqnarray}
\Delta\left( {1 \over \mu} \right) B_r^2 & - & \Delta \mu H_\theta^2 
\nonumber \\
& = & 8 {\mu - \mu' \over \mu + \mu'} \left( {I B_0 \over c a} \sin\theta
- {B_0 H_1 \over 2} \cos\theta \right) + ...
\label{s111}
\end{eqnarray}
Using this in eq.~(\ref{s108}), we again obtain the result (\ref{p1}).

\section{Appendix: The Field {\bf H} for the Example}

The magnetic field {\bf H} for the example of sec.~II is, in both cylindrical and rectangular
coordinates,
\begin{eqnarray}
{\bf H} & = & \left\{  \begin{array}{ll}
{2 \mu \over \mu + \mu'} \left( H_0 \cos\theta + H_1 {r \over a} \cos 2\theta \right) 
\hat{\bf r} 
\\
+ \left[ {2 I r \over c a^2}
- {2 \mu \over \mu + \mu'} \left( H_0 \sin\theta + H_1 {r \over a} \sin 2\theta \right) 
\right] \hat\theta  & (r < a), 
\\
\left[ H_0  
\left( 1 - {\mu - \mu' \over \mu + \mu'} {a^2 \over r^2} \right) \cos\theta 
\right.
\\
\left. \qquad
+\ H_1 \left( {r \over a} - {\mu - \mu' \over \mu + \mu'} {a^3 \over r^3} \right) \cos 2\theta 
\right] \hat{\bf r} 
& \\ 
+ \left[ {2 I \over c r}
- H_0 \left( 1 + {\mu - \mu' \over \mu + \mu'} {a^2 \over r^2} \right) \sin\theta 
\right.
\\
\left. \qquad
-\ H_1 \left( {r \over a} + {\mu - \mu' \over \mu + \mu'} {a^3 \over r^3} \right) \sin 2\theta 
\right] \hat\theta
& (r > a), 
\end{array} 
\right.
\label{s15} \\
& = & \left\{ \begin{array}{ll}
\left[ - {2 I r \over c a^2} \sin\theta
+ {2 \mu \over \mu + \mu'} \left( H_0 + H_1 {r \over a} \cos\theta \right)
 \right] \hat{\bf x} 
\\
\qquad
+\ \left( {2 I r \over c a^2} \cos\theta - {2 \mu \over \mu + \mu'} H_1 {r \over a} 
\sin\theta \right) \hat{\bf y}  & (r < a), 
\\
\left[ - {2 I \over c r} \sin\theta
+ H_0 \left( 1 - {\mu - \mu' \over \mu + \mu'} {a^2 \over r^2} \cos 2\theta \right) 
\right.
\\
\qquad \left.
+\ H_1 \left( {r \over a} \cos\theta 
- {\mu - \mu' \over \mu + \mu'} {a^3 \over r^3} \cos 3\theta \right)
 \right] \hat{\bf x}
\\ 
+ \left[ {2 I \over c r} \cos\theta
- {\mu - \mu' \over \mu + \mu'} H_0 {a^2 \over r^2} \sin 2\theta 
\right.
\\
\qquad \left.
-\ H_1 \left( {r \over a} \sin\theta 
+ {\mu - \mu' \over \mu + \mu'} {a^3 \over r^3} \sin 3\theta \right) 
\right] \hat{\bf y} & (r > a).
\end{array} \right.
\label{s16}
\end{eqnarray}


\vspace{-0.2in}

\begin{thebibliography}{[99]}

\vspace{-0.7in}

\bibitem{Casperson}
L.W.~Casperson,
{\sl Forces on permeable conductors in magnetic fields},
Am.\ J.\ Phys.\ {\bf 70}, 163-168 (2002).

\bibitem{Stratton}
See, for example, secs.~4.20-21 of J.A.~Stratton,
{\em Electromagnetic Theory}
(McGraw-Hill, New York, 1941).

\bibitem{Landau}
L.D.~Landau, E.M.~Lifshitz and L.P.~Pitaevskii,
{\em Electrodynamics of Continuous Media}, 2nd ed.,
(Butterworth-Heinemann, Oxford, 1984).

\bibitem{Brown1}
W.F.~Brown, Jr. \etal,
{\sl The teaching of electricity and magnetism at the college level. I. Logical 
standards and critical issues},
Am.\ J.\ Phys.\ {\bf 18}, 1-25 (1950).

\bibitem{Brown2}
W.F.~Brown, Jr. \etal,
{\sl The teaching of electricity and magnetism at the college level. II. Two outlines
for teachers},
Am.\ J.\ Phys.\ {\bf 18}, 69-88 (1950).

\bibitem{Brown3}
W.F.~Brown, Jr.,
{\sl Electric and magnetic forces: A direct calculation. I},
Am.\ J.\ Phys.\ {\bf 19}, 290-304 (1951).

\bibitem{Brown4}
W.F.~Brown, Jr.,
{\sl Electric and magnetic forces: A direct calculation. II},
Am.\ J.\ Phys.\ {\bf 19}, 333-350 (1951).

\bibitem{Oersted}
H.C.~Oersted,
{\sl Experimenta circa effectum conflictus electrici in acum magneticam},
(Copenhagen, 1820); Ann.\ Phil.\ {\bf 16}, 271 (1820).

\bibitem{Biot}
J.-B.~Biot and F.~Savart,
Annales de Chimie {\bf 15}, 222 (1820); J.\ de Phys.\ {\bf 41}, 151 (1820).
The modern form of the Biot-Savart force law is due to H.~Grassmann,
{\sl Neue Theorie der Elektrodynamik},
Ann.\ d.\ Phys.\ u.\ Chem.\ {\bf 64}, 1-18 (1845).

\bibitem{Ampere}
A.-M.~Amp\`ere,
{\sl La d\'etermination de la formule qui repres\'ente l'action mutuelle de deux
portions infiniment petites de conducteur Volta\"{\i}ques},
L'acad\'emie Royale des Sciences (Paris, 1822);
{\em Th\'eorie math\'ematique des ph\'enomenes \'electro-dynamiques, 
uniquement d\'eduite de l'exp\'erience},
(A.~Blanchard, Paris, 1958).

\bibitem{Poisson}
S.-D.~Poisson,
Mem.\ d.\ l'Acad.\ {\bf V}, 247 (1824).

\bibitem{permeable-wire}
K.T.~McDonald
{\sl Magnetic Force on a Permeable Wire}
(March 17, 2002), \hfill\break
http://puhep1.princeton.edu/\~mcdonald/examples/
 \hfill\break
permeable\_wire.pdf

\bibitem{Lowes}
F.J.~Lowes,
{\sl Force on a Wire in a Magnetic Field},
Nature {\bf 246}, 208-209 (1973).

\bibitem{McKinnon}
W.R.~McKinnon, S.P.~McAlister and C.M.~Hurd,
{\sl Origin of the force on a current-carrying wire in a magnetic field},
Am.\ J.\ Phys.\ {\bf 49}, 493-494 (1981).

\bibitem{Panofsky}
See, for example, sec.~1.1 of W.K.H.~Panofsky and M.~Phillips,
{\em Classical Electricity and Magnetism}, 2nd ed.\
(Addison-Wesley, Reading, MA, 1962).

\bibitem{Page}
See, for example, L.~Page and N.I.~Adams,
{\sl Action and Reaction Between Moving Charges},
Am.\ J.\ Phys.\ {\bf 13}, 141-147 (1945).

\bibitem{Cavalleri}
G.~Cavalleri, G.~Spavieri and G.~Spinelli,
{\sl The Amp\`ere and Biot-Savart force laws},
Eur.\ J.\ Phys.\ {\bf 17}, 205-207 (1996).

\bibitem{Rosensweig}
R.E.~Rosensweig,
{\em Ferrohydrodynamics}
(Constable, London, 1985; Dover Publications, New York, 1997).

\bibitem{Lorentz}
H.A.~Lorentz,
{\sl Versuch einer Theorie der electrischen und optischen Ersheinungen in bewegten 
K\"orpern}
(E.J.\ Brill, Leiden, 1895), Art. 12.

\bibitem{Jefimenko}
O.D.~Jefimenko,
{\em Electricity and Magnetism}, 2nd ed.\
(Electret Scientific Co., Star City, 1989).

\bibitem{Thomson}
See p.~499 of W.~Thomson, 
{\em Papers on Electrostatics and Magnetism} 
(Macmillan, London, 1884).

\bibitem{Rasetti}
F.~Rasetti,
{\sl Deflection of mesons in magnetized iron},
Phys.\ Rev.\ {\bf 66}, 1-5 (1944).  This paper mentions the earlier history of
erratic results on this topic.

\bibitem{Corson}
P.~Lorrain, D.R.~Corson and F.~Lorrain,
{\em Electromagnetic Fields and Waves}, 3rd ed.\
(W.H.~Freeman, New York, 1988).

\bibitem{BorH}
Poisson worked exclusively with the magnetic field {\bf H}, but realized that the
force on a fictitious magnetic pole $p$ is not necessarily ${\bf F} = p {\bf H}$,
since poles are always found inside bulk media, which results in an additional
force on the pole depending on the assumed shape of the surrounding cavity.
W.~Thomson (Lord Kelvin) noted in 1850 \cite{Thomson}
that for a pole in a disk-shaped cavity 
with axis parallel to the magnetization {\bf M} of the medium, 
the force would be ${\bf F} = p ({\bf H} + 4 \pi {\bf M})$,
and therefore he introduced the magnetic field ${\bf B} = {\bf H} + 4 \pi {\bf M}$
``according to the electromagnetic definition''.
  In sec.~400 of his {\em Treatise} \cite{Maxwell}, Maxwell follows
Thomson in stating that the force on a magnetic pole is usefully considered to be
${\bf F} = p {\bf B}$.  This is very reasonable since in linear magnetic media
fictitious poles are only found on surfaces, for which an appropriate
 surrounding cavity is disk-shaped.

\bibitem{Maxwell}
J.C.~Maxwell,
{\em A Treatise on Electricity and Magnetism}, 3rd. ed.\
(Clarendon Press, Oxford, 1891; Dover Publications, New York, 1954).

\bibitem{Jackson}
J.D.~Jackson,
{\em Classical Electrodynamics}, 3rd ed.\
(Wiley, New York, 1999).

\bibitem{Helmholtz}
H.~von Helmholtz,
{\sl \"Uber die auf das Innere magnetisch order dielectrisch polarisirter K\"orper
wirkenden Kr\"afte},
Ann.\ d.\ Phys.\ {\bf 13}, 385-406 (1882).

\bibitem{Brevik}
For a pedagogic review of electrostriction and magnetostriction, see I.~Brevik,
{\sl Fluids in electric and magnetic fields: Pressure variation and stability},
Can.\ J.\ Phys.\ {\bf 60}, 449-455 (1982).  More extensive discussion is given 
in \cite{Rosensweig}.

\bibitem{Maxwell641}
See secs.~641-645 of \cite{Maxwell}.

\bibitem{Maxwell501}
See sec.~501 of \cite{Maxwell}.

\end{thebibliography}
\end{document}